\documentclass[a4paper,12pt]{article}
\usepackage{amsmath}
\usepackage{amssymb}
\usepackage{amsfonts}
\usepackage{graphicx}
\usepackage{enumitem}
\usepackage{natbib}
\usepackage{authblk}
\usepackage{xr}
\usepackage[toc,page]{appendix}
\usepackage{enumitem}
\graphicspath{{./figure/}}
\renewcommand{\mathbf}[1]{\boldsymbol{#1}}

\begin{document}

\title{A Robust Seemingly Unrelated Regressions For Row-Wise And Cell-Wise Contamination}

\author[1]{Giovanni Saraceno}
\author[2]{Fatemah Alqallaf}
\author[1]{Claudio Agostinelli}

\affil[1]{Department of Mathematics, University of Trento, Trento, Italy}
\affil[2]{Department of Statistics and Operations Research, Kuwait University, Kuwait}

\date{\today}
\maketitle

\begin{abstract}
The Seemingly Unrelated Regressions (SUR) model is a wide used estimation procedure in econometrics, insurance and finance, where very often, the regression model contains more than one equation. Unknown parameters, regression coefficients and covariances among the errors terms, are estimated using algorithms based on Generalized Least Squares or Maximum Likelihood, and the method, as a whole, is very sensitive to outliers. To overcome this problem M-estimators and S-estimators are proposed in the literature together with fast algorithms. However, these procedures are only able to cope with row-wise outliers in the error terms, while their performance becomes very poor in the presence of cell-wise outliers and as the number of equations increases. A new robust approach is proposed which is able to perform well under both contamination types as well as it is fast to compute. Illustrations based on Monte Carlo simulations and a real data example are provided.
\end{abstract}

\smallskip

\section{Introduction} 
\label{SecIntroduction}
The Seemingly Unrelated Regression (SUR) model or  Seemingly Unrelated Regression Equations (SURE), proposed by \citet{zellner1962}, is a generalization of a linear regression model that consists of several regression equations, each having its own dependent variable and potentially different set of exogenous explanatory variables. Each equation is a valid linear regression on its own and can be estimated separately, however the errors are assumed to be correlated across the equations.

The model can be estimated equation-by-equation using standard ordinary least squares (OLS). Such estimates are consistent, however generally not as efficient as the SUR method, which amounts to feasible generalized least squares (FGLS) \citep{zellner1962} with a specific form of the variance-covariance matrix. The SUR model is equivalent to OLS method in two particular cases: when the errors are uncorrelated between the equations, so that they are truly unrelated, and when each equation contains exactly the same set of regressors. Furthermore, it can be viewed as either the simplification of the general linear model where certain coefficients are restricted to be equal to zero, or as the generalization of the general linear model where the regressors on the right hand side are allowed to be different in each equation. Finally, the SUR model can be further generalized into the simultaneous equations model, where the regressors are allowed to be the endogenous variables as well. Other estimation procedures besides FGLS have been proposed for SUR models, see \cite{kmenta1968}.

However, all these proposed methods are not robust, since they are based on least squares and outliers, defined as observations separated from the bulk of data, can highly affect the estimation procedure. Hence, robust alternatives have been proposed. \cite{koenker1990} introduced a robust procedure for the SUR model based on M-estimation, but it is not affine equivariant. Then, \cite{rousseeuw1984} and \cite{lopuha1989} studied S-estimators for multivariate location and scatter, whereas \cite{bilodeau2000} firstly introduced S-estimation in regression problems. The robust SUR estimator proposed by \cite{bilodeau2000} results to be regression and affine invariant, and has nice robust properties, such as high breakdown point, but it is computationally expensive. To overcome the computational issue, \cite{hubert2017} introduced a fast and robust SUR method based on Fast-S algorithm \citep{salibian2006a,salibian2006b}.

The robust estimators of SUR model seen so far deal with row-wise contamination, Tukey-Huber contamination model (THCM), which assumes that a proportion $\epsilon$ of observations can be contaminated and these independent events are the units considered as outliers. \cite{alqallaf2009} consider a different contamination model for multivariate data: the independent contamination model (ICM) or cell-wise contamination, where the entries of an observation (or cells) can be independently contaminated. According to this paradigm, given a fraction $\epsilon$ of contaminated cells, the probability that at least one component of an observation is an outlier is $1 - (1-\epsilon)^p$, where $p$ is the dimension of observations. This number is close to one when $p$ is large even if $\epsilon$ is small. For this reason estimators that have breakdown point 0.5 under the THCM, may have breakdown tending to zero under the ICM. \cite{alqallaf2009} show that this happens with the most popular high breakdown point equivariant estimators of multivariate location, e.g., S-estimators\citep{davies1987}, Minimum Volume Ellipsoid \citep{rousseeuw1985}, Minimum Covariance Determinant \citep{rousseeuw1985} or the Stahel-Donoho estimators \citep{donoho1982, stahel1981}. In order to cope with both row-wise and cell-wise contamination, \cite{agostinelli2015} developed the 2SGS-estimators for multivariate location and scatter while \cite{leung2017} further extended the approach to linear models.

Here, we propose a robust estimator for the SUR model, which we will refer to as \textit{surerob}, able to deal with both row-wise contamination and cell-wise contamination by robustfying the FGLS approach of \cite{zellner1962}. To the best of our knowledge, there are no competiting estimators in literature which are able to deal with cell-wise outliers for SUR models.

The remainder of the paper is organized as follows. The SUR model is presented in Section \ref{sec:sur}, while Section \ref{sec:estimation} describes the proposed robust estimation method. The performance of the estimators, in case of row-wise and cell-wise contamination, is investigated through a simulation study in Section \ref{sec:simulation} and with a real data example in Section \ref{sec:example}. Concluding remarks end the paper in Section \ref{sec:conclusions}. 

\section{The SUR model}
\label{sec:sur}

Consider $m$ regression equations
\begin{equation*}
y_{ik} = \mathbf{x}_{ik}^\top \boldsymbol{\beta}_i + \varepsilon_{ik} \ , \qquad i=1, \ldots, m \ .
\end{equation*}
Here, $i$ represents the equation index, while $k=1,\ldots,n$ is the observation index and $n$ is the total number of observations. Each equation $i$ has a single response variable $\mathbf{y}_i = (y_{i1}, \ldots, y_{in})^\top$, errors vector $\boldsymbol{\varepsilon}_i = (\varepsilon_{i1}, \ldots, \varepsilon_{in})^\top$, an explanatory $n \times p_i$ matrix $X_i$ and $p_i$-vector $\boldsymbol{\beta}_i$ of coefficients which lead to the form
\begin{equation*}
\mathbf{y}_i = X_i \boldsymbol{\beta}_i + \boldsymbol{\varepsilon}_i \ , \qquad i=1,\ldots, m \ .
\end{equation*}
Finally, if we stack these $m$ vector equations on top of each other, the system will take the form \citep[][eq. (2.2)]{zellner1962}
\begin{align} \label{equ:sur:univariate}
\begin{pmatrix} \mathbf{y}_1 \\ \mathbf{y}_2 \\ \vdots \\ \mathbf{y}_m \end{pmatrix} & = 
    \begin{pmatrix} X_1 & 0 & \ldots & 0 \\ 0 & X_2 & \ldots & 0 \\ \vdots & \vdots & \ddots & \vdots \\ 0 & 0 & \ldots & X_m \end{pmatrix}
    \begin{pmatrix} \boldsymbol{\beta}_1 \\ \boldsymbol{\beta}_2 \\ \vdots \\ \boldsymbol{\beta}_m \end{pmatrix} +
    \begin{pmatrix} \boldsymbol{\varepsilon}_1 \\ \boldsymbol{\varepsilon}_2 \\ \vdots \\ \boldsymbol{\varepsilon}_m \end{pmatrix} \\ \nonumber
\mathbf{y} & = X \boldsymbol{\beta} + \boldsymbol{\varepsilon} \ .
\end{align}
where $\boldsymbol{\beta}$ is a vector of dimension $P = \sum_{i=1}^m p_i$. The assumption of the model is that error terms $\varepsilon_{ik}$ are independent across observations, but may have cross-equation contemporaneous correlations, that is, $\mathbb{E}( \varepsilon_{ir} \varepsilon_{is} | X ) = 0$ whenever $r \neq s$, whereas $\mathbb{E}(\varepsilon_{ik} \varepsilon_{jk} | X) = \sigma_{ij}$. Note that $\sigma_{ii}$ is the variance of the error term in the $i$th equation, whereas $\sigma_{ij}$ is the covariance between the errors in equations $i$ and $j$. Let $\Sigma$ denote the $m\times m$ covariance matrix of each observation with entries $\sigma_{ij}$, then the covariance matrix of the stacked error terms $\boldsymbol{\varepsilon}$ will be equal to 
\begin{equation*}
\Omega = \mathbb{E} ( \boldsymbol{\varepsilon} \boldsymbol{\varepsilon}^\top | X ) = \Sigma \otimes I_n,
\end{equation*}
where $I_n$ is the $n \times n$ identity matrix and $\otimes$ denotes the matrix Kronecker product.

An alternative formulation of the multivariate SUR model is given by
\begin{equation} \label{equ:sur:multivariate}
\tilde{Y} = \tilde{X} B + E
\end{equation}
where $\tilde{Y} = (\mathbf{y}_1, \ldots, \mathbf{y}_m)$ is the $n \times m$ response matrix, $\tilde{X} = (X_{1}, \ldots, X_{m})$ is the $n \times P$ (recall that $P=\sum_{i=1}^m p_i$) design matrix, $B = \operatorname{diag}(\boldsymbol{\beta}_1, \ldots, \boldsymbol{\beta}_m)$ is a $P \times m$ block diagonal matrix, and $E = (\boldsymbol{\varepsilon}_1, \ldots, \boldsymbol{\varepsilon}_m)$ with $\operatorname{\mathbb{C}ov}(E) = \Omega = \Sigma \otimes I_n$. 

Let $\mathbf{T} = (\mathbf{t}_{1},\ldots,\mathbf{t}_{n})^\top$ be a data set of size $n$ corresponding to model (\ref{equ:sur:multivariate}), where $\mathbf{t}_{k}= (\mathbf{t}_{1k}, \ldots, \mathbf{t}_{mk})$ and $\mathbf{t}_{ik}=(y_{ik},x_{ik1},\ldots,x_{ikp_i})$, $1 \leq i \leq m$, $1 \le k \le n$, and $x_{ikp}$ is the value in the $k$-th row and $p$-th column of the matrix $ X_i$. According to THCM a proportion $\epsilon$ of observations $\mathbf{t}_{k}$, $1 \leq k \leq n$, can be contaminated, while following ICM the entries  $\mathbf{t}_{ik}$, $1 \leq k \leq n$, $1 \leq i \leq m$ can be independently replaced by an outlier.

\section{Robust estimation}
\label{sec:estimation}

We discuss the two main estimation methods used for the SUR model: the standard FGLS method \citep{zellner1962} and the robust SUR method base on Fast-S algorithm \citep{hubert2017}, which we will refer to as \textit{fastSUR}. Finally, we introduce the proposed new estimator for the SUR model robust against row-wise and cell-wise outliers.

The SUR model is usually estimated using  FGLS method \citep{zellner1962}. This is a two-step method where in the first step an ordinary least squares regression is performed for each model equation separately. The residuals $\hat{\boldsymbol{\varepsilon}_i}$ from these regressions are used to estimate the elements of the matrix $\Sigma$ by computing the sample covariance matrix $\hat{\Sigma}_1$ with components
\begin{equation}
\hat{\sigma}_{ij} = \frac{1}{n} \hat{\boldsymbol{\varepsilon}}_i^\top \hat{\boldsymbol{\varepsilon}}_j \qquad i,j=1,\ldots,m \ .
\end{equation}
In the second step a generalized least squares regression is performed using the variance matrix $\hat\Omega = \hat\Sigma_1 \otimes I_n$ to obtain final estimates of the regression coefficients as
\begin{equation}
\hat{\boldsymbol{\beta}} = \left( X^\top \hat{\Omega} X \right)^{-1} X^\top \hat\Omega^{-1} \mathbf{y} \ .
\end{equation}
Residuals can then be recomputed and used to obtain a final estimate $\hat{\Sigma}_2$ of the covariance matrix  $\Sigma$.
The FGLS algorithm is available in R \citep{Rcore} in package \texttt{systemfit}, see \citet{henningsen2007}.

The fastSUR, which is the computationally efficient version of the robust SUR method of \cite{bilodeau2000}, is the couple $(\hat{B}, \hat{\Sigma})$ so that
\begin{equation*}
(\hat{B}, \hat{\Sigma}) = \arg\min_{(B,S)} |S|   
\end{equation*}
under the condition
\begin{equation*}
\frac{1}{n} \sum_{k=1}^n \rho\left( \mathbf{e}_i(B)^\top S^{-1} \mathbf{e}_i(B) \right) = b
\end{equation*}
where $|\cdot|$ is the determinant, $B = \operatorname{diag}(\mathbf{\beta}_1, \ldots, \mathbf{\beta}_m)$, $\mathbf{\beta}_i \in \mathbb{R}^{p_i}$ ($i=1, \ldots, m$), $\mathbf{e}_i(B) = \tilde{Y} - \tilde{X} B$, and $S$ is an $m \times m$ symmetric positive definite matrix. The function $\rho$ \citep[see, e.g., ][]{maronna2018} is chosen so that
\begin{itemize}
\item[(C1)] $\rho$ is symmetric around zero and twice continuously differentiable;
\item[(C2)] $\rho(0) = 0$ and $\rho$ is strictly increasing on $[0, c_0]$ and constant on $[c_0, \infty]$ for some $c_0 > 0$.
\end{itemize}
The constant $b$ can be computed as $\mathbb{E}_{F_0}(\rho(|\mathbf{e}|))$, where $\mathbf{e} \sim F_0$ and $F_0 = N_m(\mathbf{0}, I_m)$ which ensures consistency at the model with normal errors. 
Similar to classic robust location and scatter estimators, robust SUR and its computationally efficient version fastSUR are affine equivariant and their  breakdown point tends to zero as the number of equations $m$ gets larger under the Independent Contamination Model.

We propose a robust estimator for the SUR model under both types of contamination by robustfying the FGLS approach of \citet{zellner1962}. In the first step, residuals $\hat{\boldsymbol{\varepsilon}_i}$ are estimated by means of an MM-estimator of regression, as introduced in \citet{yohai1987},  equation by equation. Since these are univariate regression models, the procedure achieves a breakdown of $0.5$ and the estimates of the regression coefficients are affine equivariant. At this point, the ICM contamination scheme is considered, therefore we construct the residual data matrix $\hat{E} = (\hat{\boldsymbol{\varepsilon}}_1, \ldots, \hat{\boldsymbol{\varepsilon}}_m)$ where each cell of this matrix could be a cell-wise outlier. Let $W_i = \operatorname{diag}(\mathbf{w}_i)$ be an $n \times n$ diagonal matrix where $\mathbf{w}_i$ is the vector of robust weights associated to each observation in the $i$th equation, that is, the $k$th element $w_{ik}$ of the vector $\mathbf{w}_i$ is given by 
  \begin{equation}
    \label{eq:weights}
w_{ik} = \left\{
\begin{array}{ll}
\psi(\hat{\varepsilon}_{ik}/s_i) / (\hat{\varepsilon}_{ik}/s_i) & \text{if} \ \hat{\varepsilon}_{ik} \neq 0 \\
1 & \text{otherwise}
\end{array}
\right. 
\end{equation}
where $\psi = \rho'$ is the first derivative of the function $\rho$ and $s_i$ is the estimated standard deviation of the errors for the $i$th equation. A robust estimate $\hat{\Sigma}_1$ of the covariance matrix $\Sigma$ is obtained using the 2SGS method based on the residual data matrix $\hat{E}$. 
The final estimate of the coefficients is than obtained as
\begin{equation*}
\hat{\boldsymbol{\beta}} = \left(X^\top W (\hat{\Sigma}_1^{-1} \otimes I_n ) W X \right)^{-1} X^\top W (\hat{\Sigma}_1^{-1} \otimes I_n ) W \mathbf{y} \ ,
\end{equation*}
where $W$ is a block diagonal matrix with $(W_1, \ldots, W_m)$ in the main diagonal. New residuals can then be obtained and, applying 2SGS to them, we get the final estimate $\hat{\Sigma}_2$ of the covariance matrix.

\section{Simulation Study}
\label{sec:simulation}

The performance of the introduced method, which we will refer to as \textit{surerob}, is compared with classical FGLS algorithm, as implemented in the R package \texttt{systemfit} \citep{henningsen2007}, indicated as \textit{sure}, and the fastSUR algorithm \citep{hubert2017}. The fastSUR algorithm \citep{hubert2017} uses the Tukey's bisquare function with constant $b$ such that the breakdown point is $0.5$ \citep{rousseeuw1984} as it is implemented in an R code kindly made available by Prof. M. Hubert.

The R implementation of the surerob procedure uses the function \texttt{lmrob} in the R package \texttt{robustbase} \citep{maechler2016} and the function \texttt{TSGS} in the R package \texttt{GSE} \citep{leung2019} using default values for both. The code is available in the R package \texttt{robustsur} provided as supplementary material. An allustration of how to use the functions in the R package \texttt{robustsur} is reported in Section SM--1 of the Supplemental Material.

The simulation has the following setting: sample size $n = 100$; $p = p_i = 5$, $10$, equals for each equation and $m = 5$, $10$, $20$. For each combination of these factors we run $N = 1000$ Monte Carlo replications. The regression coefficients $\boldsymbol{\beta}_i$ are sampled from a standard Cauchy random variable for each sample while the variances in $\operatorname{diag}(\Sigma)$ are all equal to $1$, that is, $\Sigma$ is a correlation matrix. To account for the lack of affine equivariance of the proposed estimator, we consider different correlation structures. 
In particular, for each sample in our simulation we create a different random correlation matrix with condition number fixed at $\text{CN} = 100$. Correlation matrices with high condition number are less favorable for our proposed estimator. For the details about the procedure used to obtain such random correlation matrices, see \citet{agostinelli2015}.

Two types of outliers are considered: (i) generated by THCM and (ii) generated by ICM. When the outliers are generated using THCM, we randomly replace $5\%$, $10\%$, $20\%$ and $30\%$ of the cases in the errors data matrix by $k \mathbf{v}$, where $k = 0, 5, 10, \ldots, 95, 100$ and $\mathbf{v}$ is the eigenvector corresponding to the smallest eigenvalue of $\Sigma$ with length such that $\mathbf{v}^\top \Sigma^{-1} \mathbf{v} = 1$.  Monte Carlo experiments in \citet{agostinelli2015} show that the placement of outliers in this direction, $\mathbf{v}$, is the least favorable for the 2SGS estimator. When the outliers are generated using ICM, we randomly replace $5\%$, $10\%$, $20\%$ or $30\%$ of the cells in the error data matrix by the value $k$ where $k=1, 5, \dots, 95, 100$. 

The performance of a given regression coefficients estimator $\hat{\boldsymbol{\beta}}$ is measures by Mean Square Error
\begin{equation*}
\operatorname{MSE}(\hat{\boldsymbol{\beta}}, \boldsymbol{\beta}) = \frac{1}{N} \sum_{r=1}^N (\hat{\boldsymbol{\beta}}_r - \boldsymbol{\beta})^\top  (\hat{\boldsymbol{\beta}}_r - \boldsymbol{\beta})
\end{equation*}
where $\hat{\boldsymbol{\beta}}_{r}$ is the estimate at the $r$-th replication. 

The performance of a given scatter estimator $\hat{\Sigma}$ is measured by the Kullback-Leibler divergence between two Gaussian distribution with the same mean and covariances $S$ and $\Sigma$:
\begin{equation*}
\delta(S, \Sigma) = \operatorname{trace}(S \Sigma^{-1}) - \log(|S \Sigma^{-1}|) - p \ .
\end{equation*}
This divergence also appears in the likelihood ratio test statistics for testing the null hypothesis that a multivariate normal distribution has covariance matrix $\Sigma$. Then, the performance of an estimator $\hat\Sigma$ is summarized by
\begin{equation*}
\Delta(\hat{\Sigma}, \Sigma) =\frac{1}{N} \sum_{r=1}^{N} \delta(\hat{\Sigma}_r, \Sigma)
\end{equation*}
where $\hat{\Sigma}_{r}$ is the estimate at the $r$-th replication. 

\begin{figure}
\begin{center}
\includegraphics[width=0.45\textwidth]{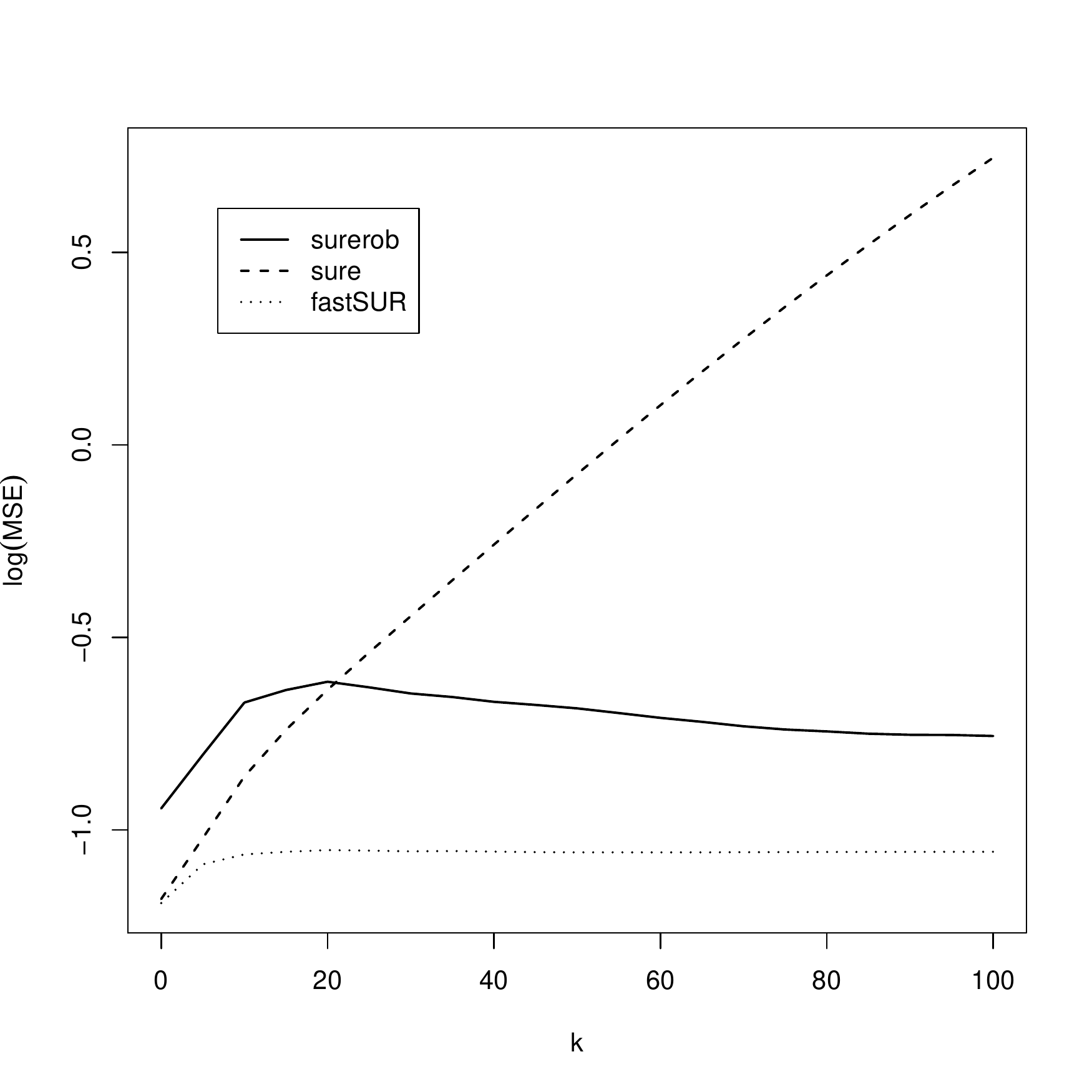}
\includegraphics[width=0.45\textwidth]{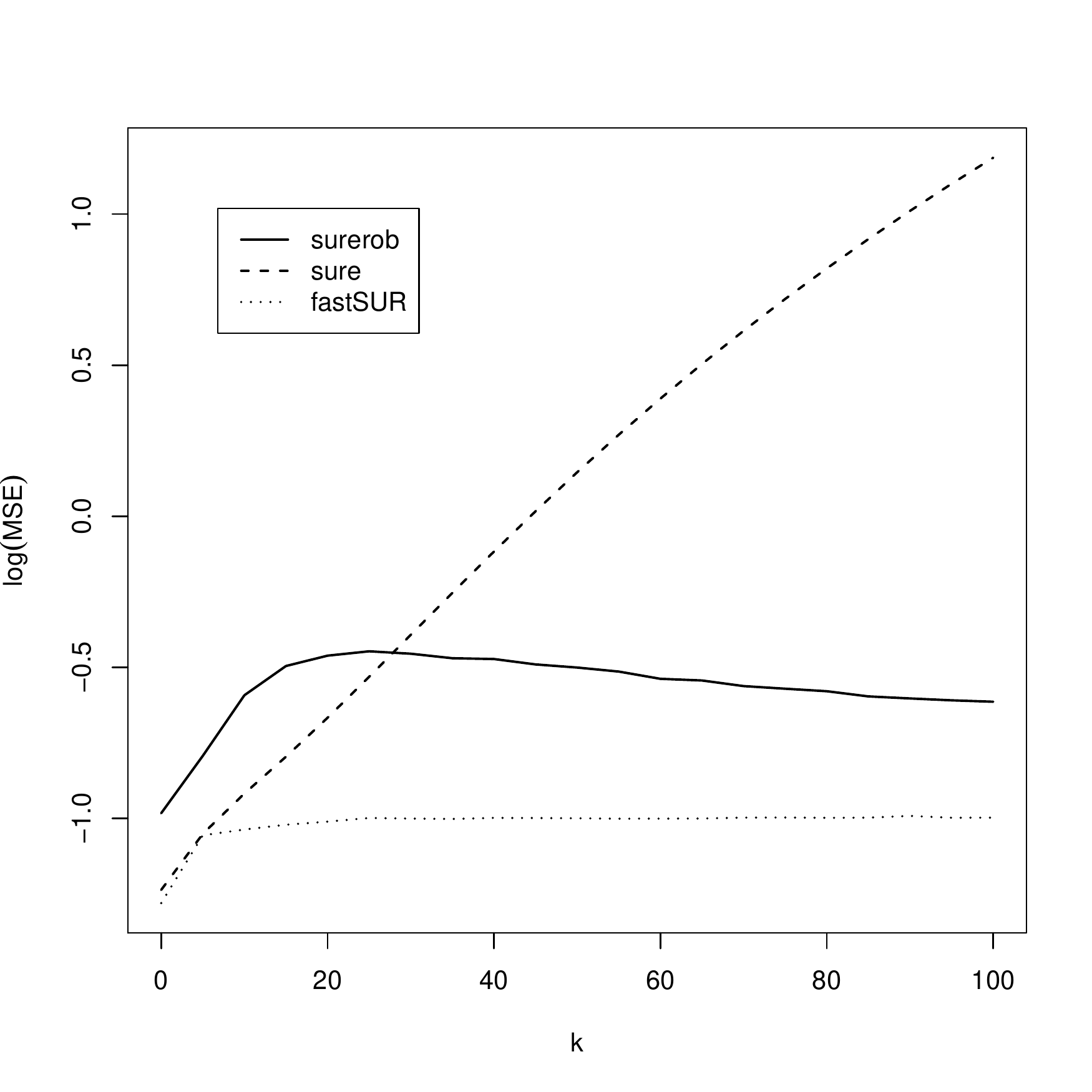} \\
\includegraphics[width=0.45\textwidth]{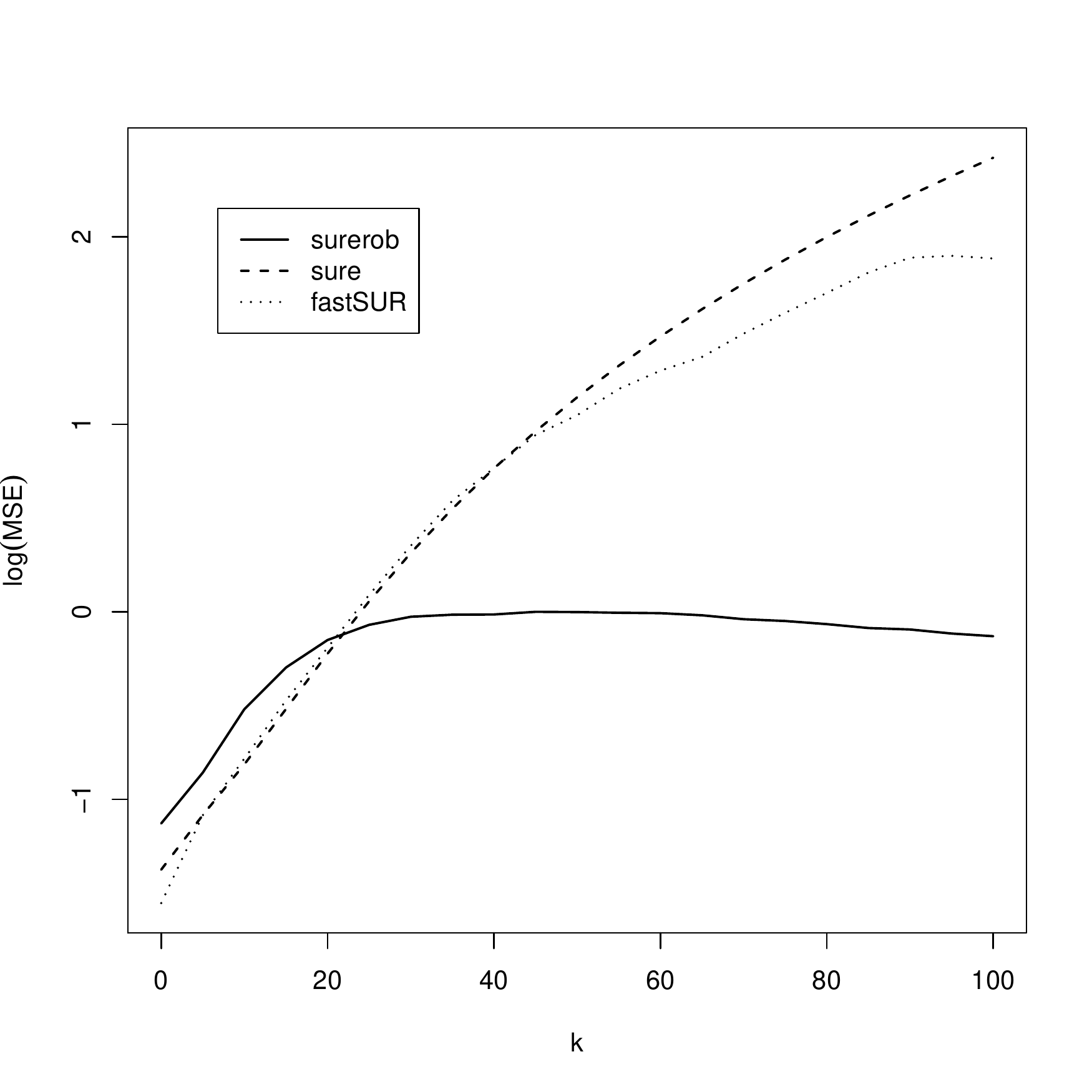}
\includegraphics[width=0.45\textwidth]{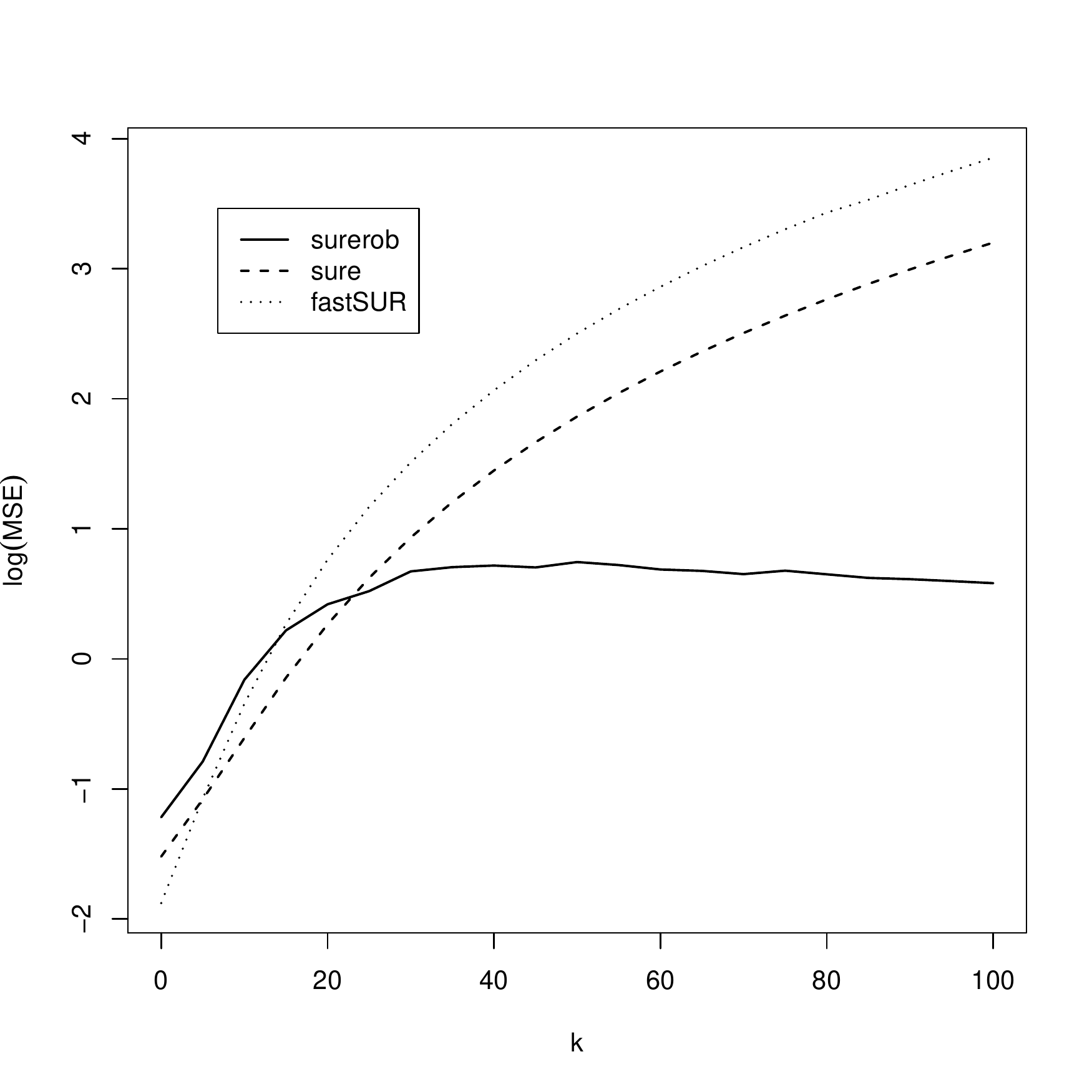}
\end{center}
\caption{Mean Square Error $\operatorname{MSE} = \operatorname{MSE}(\hat{\boldsymbol{\beta}}, \boldsymbol{\beta})$ for surerob (solid line), sure (dashed line) and fastSUR (dotted line) for different levels of contamination $\epsilon=5\%$, $10\%$, (top) $20\%$, $30\%$ (bottom) under the THCM considering dimension $p=5$ and number of equations $m=10$.}
\label{fig:mse:100:htcm}
\end{figure}

\begin{figure}
\begin{center}
\includegraphics[width=0.45\textwidth]{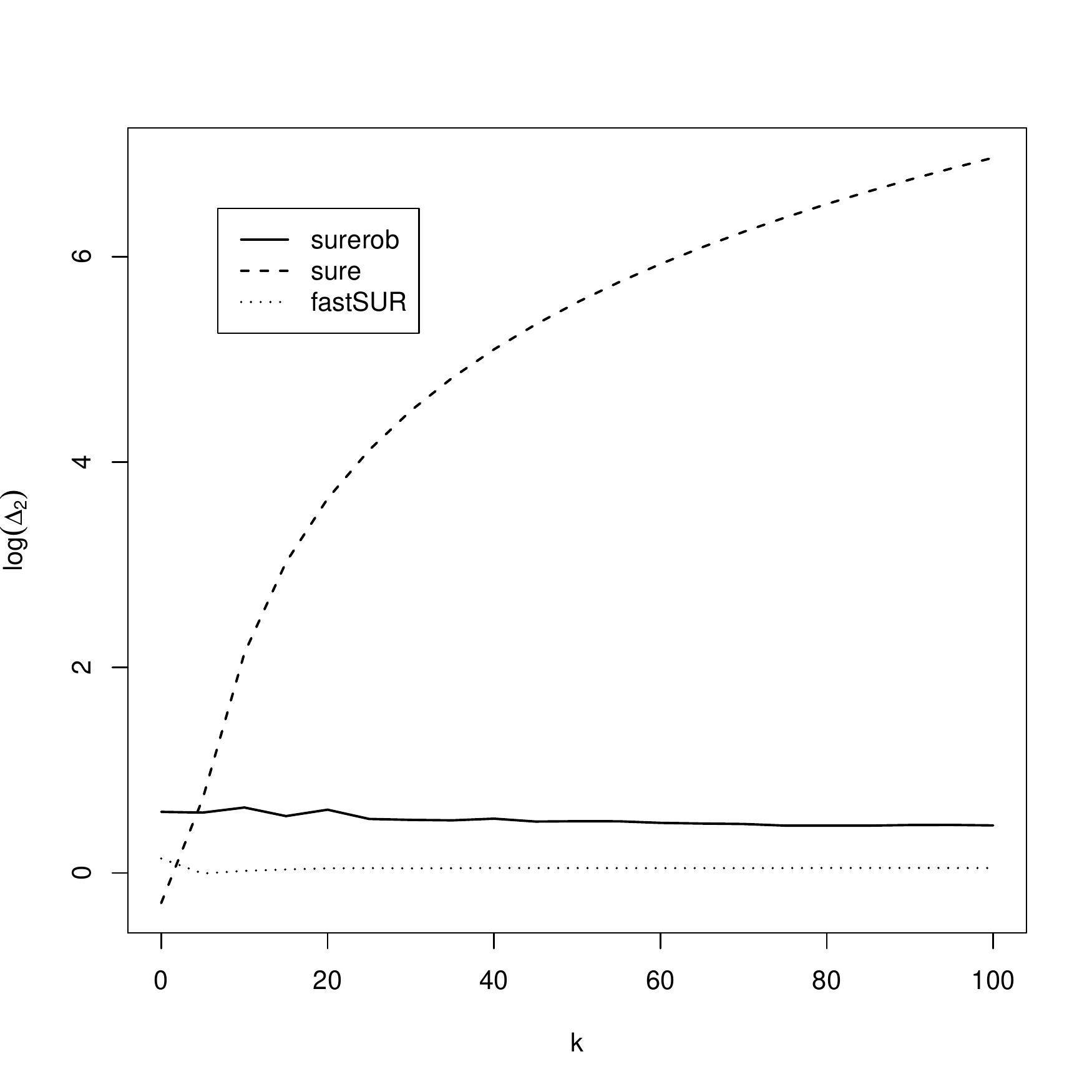}
\includegraphics[width=0.45\textwidth]{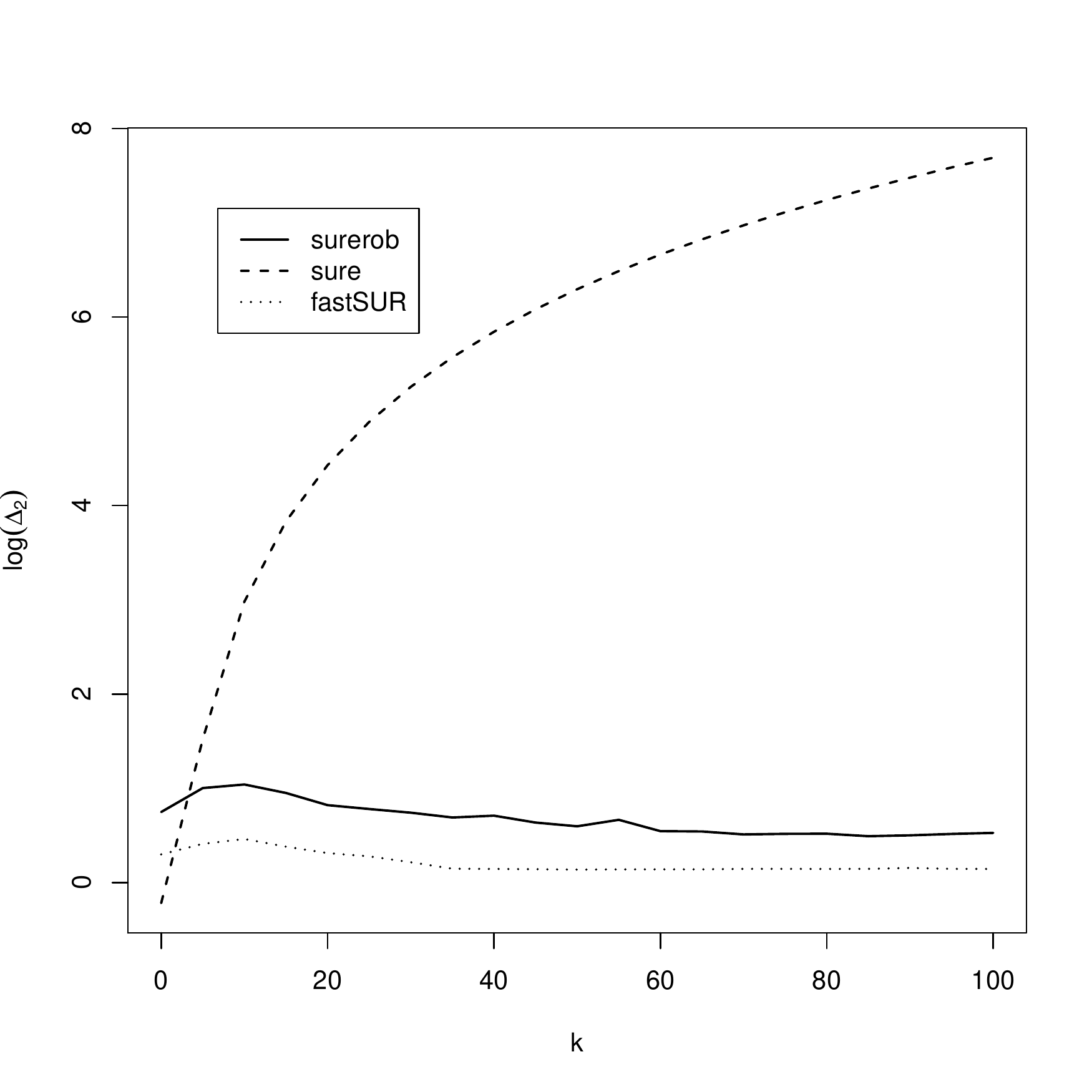} \\
\includegraphics[width=0.45\textwidth]{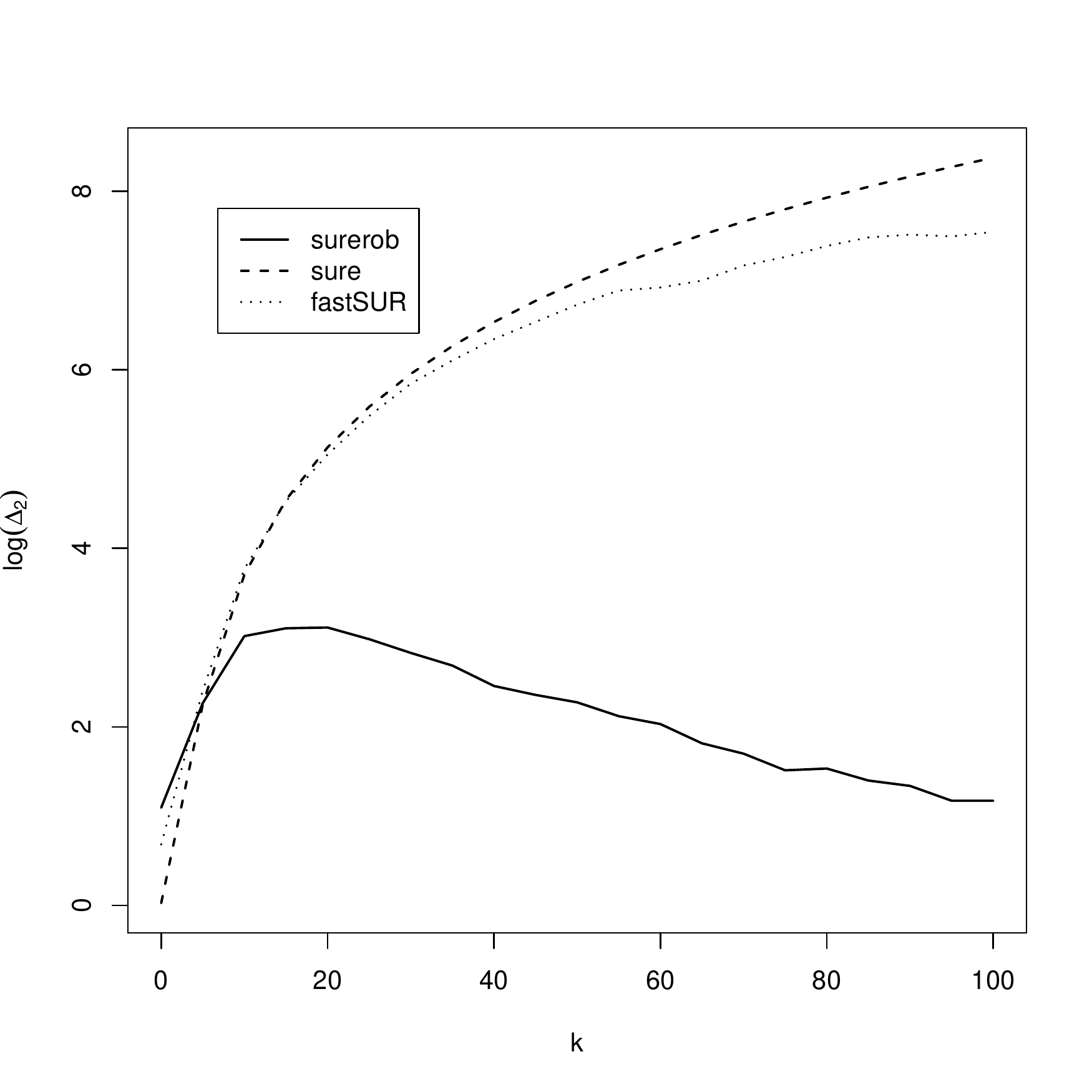}
\includegraphics[width=0.45\textwidth]{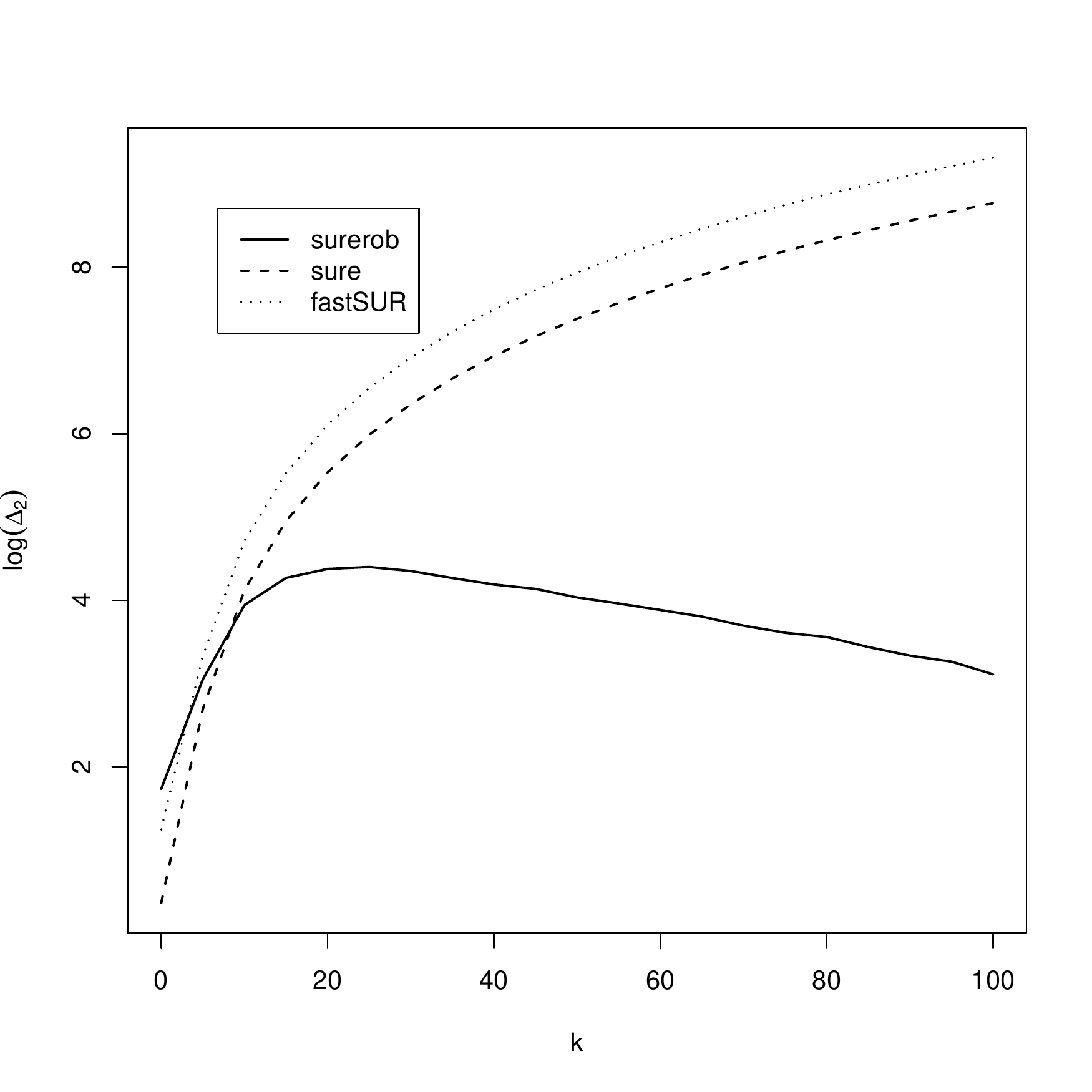}
\end{center}
\caption{Average Kullback-Leibler $\Delta_2 = \Delta(\hat{\Sigma}_2, \Sigma)$ for surerob (solid line), sure (dashed line) and fastSUR (dotted line) for different levels of contamination $\epsilon=5\%$, $10\%$, (top) $20\%$, $30\%$ (bottom) under the THCM considering dimension $p=5$ and number of equations $m=10$.}
\label{fig:div2:100:htcm}
\end{figure}

\begin{figure}
\begin{center}
\includegraphics[width=0.45\textwidth]{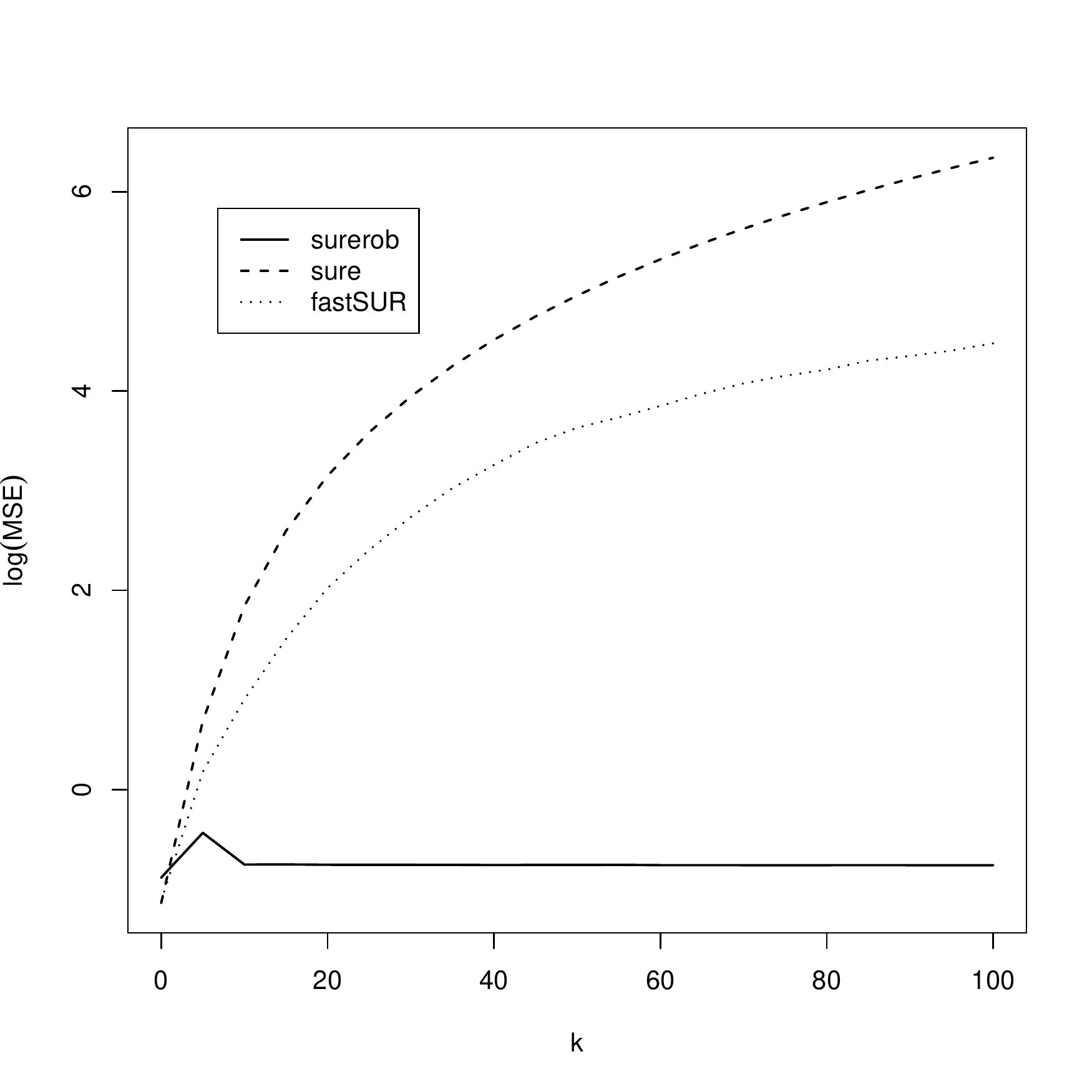}
\includegraphics[width=0.45\textwidth]{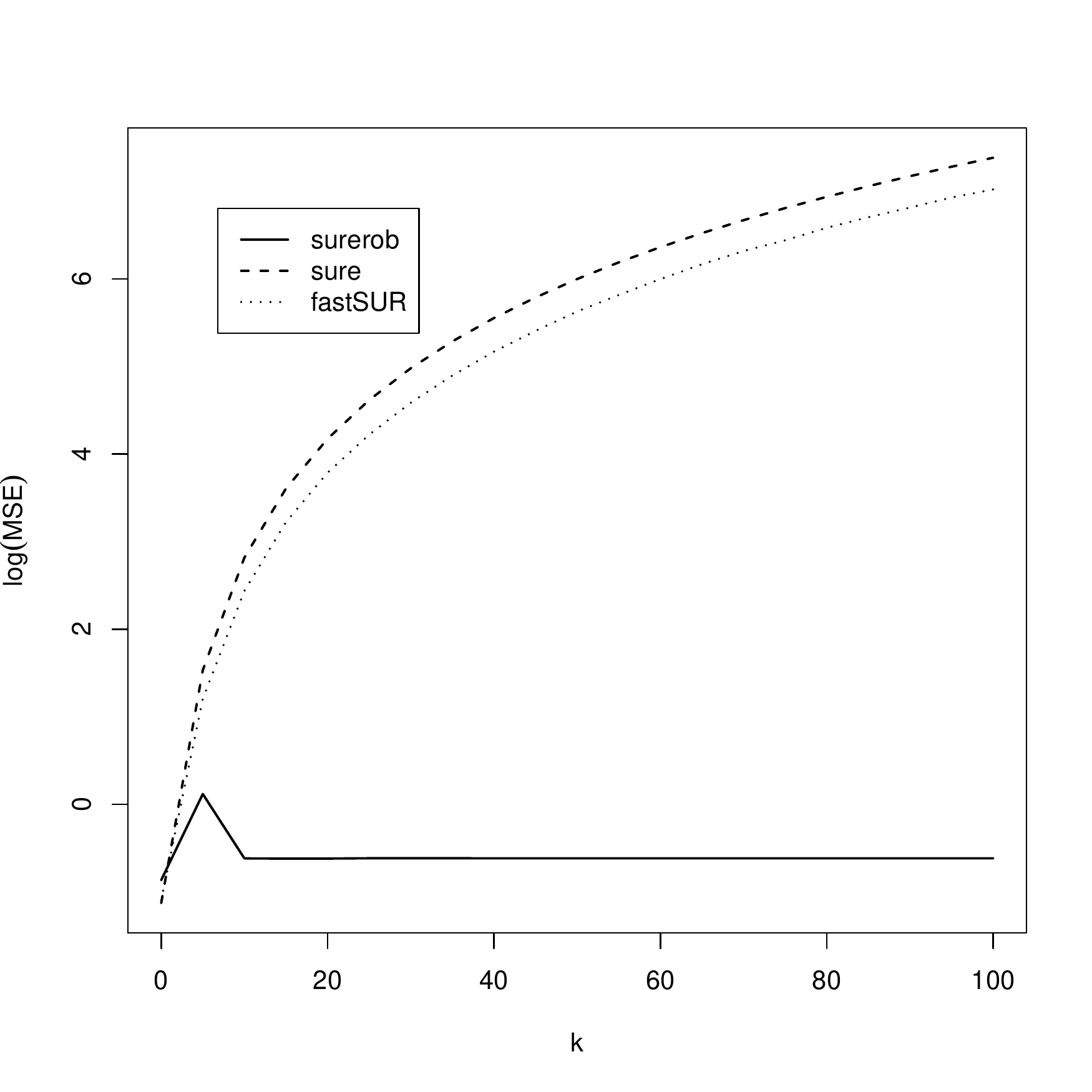} \\
\includegraphics[width=0.45\textwidth]{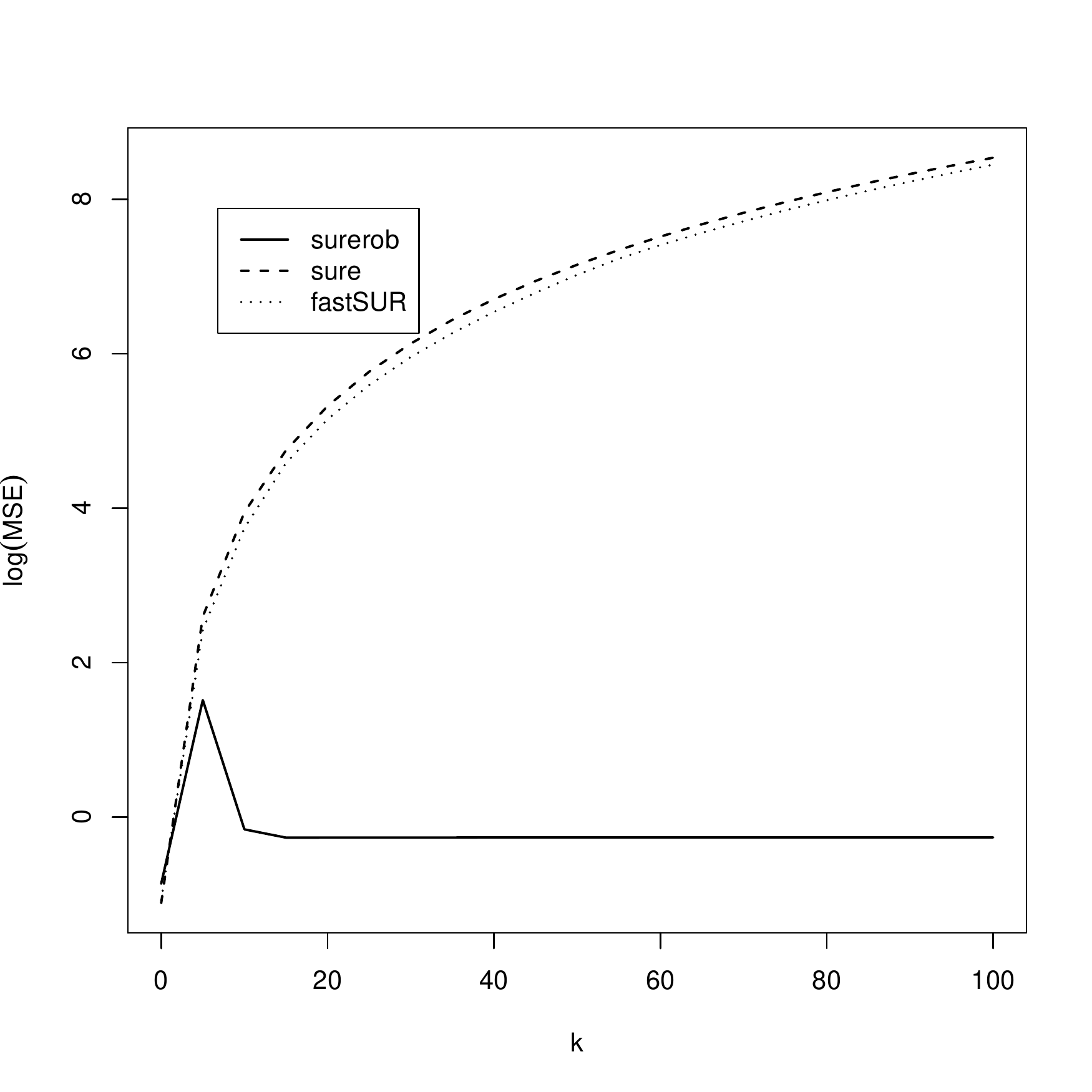}
\includegraphics[width=0.45\textwidth]{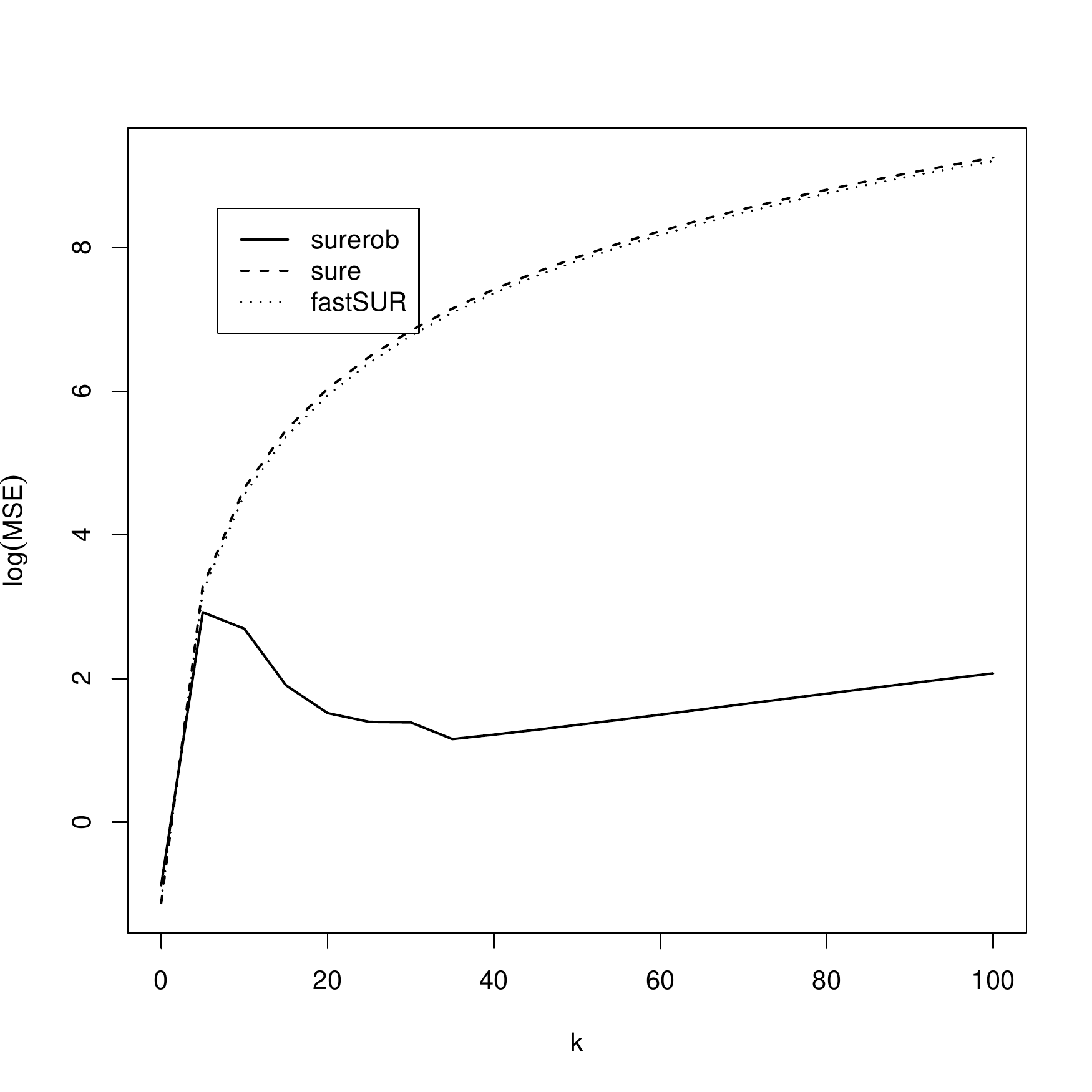}
\end{center}
\caption{Mean Square Error $\operatorname{MSE} = \operatorname{MSE}(\hat{\boldsymbol{\beta}}, \boldsymbol{\beta})$ for surerob (solid line), sure (dashed line) and fastSUR (dotted line) for different levels of contamination $\epsilon=5\%$, $10\%$, (top) $20\%$, $30\%$ (bottom) under the ICM considering dimension $p=5$ and number of equations $m=10$.}
\label{fig:mse:100:icm}
\end{figure}

\begin{figure}
\begin{center}
\includegraphics[width=0.45\textwidth]{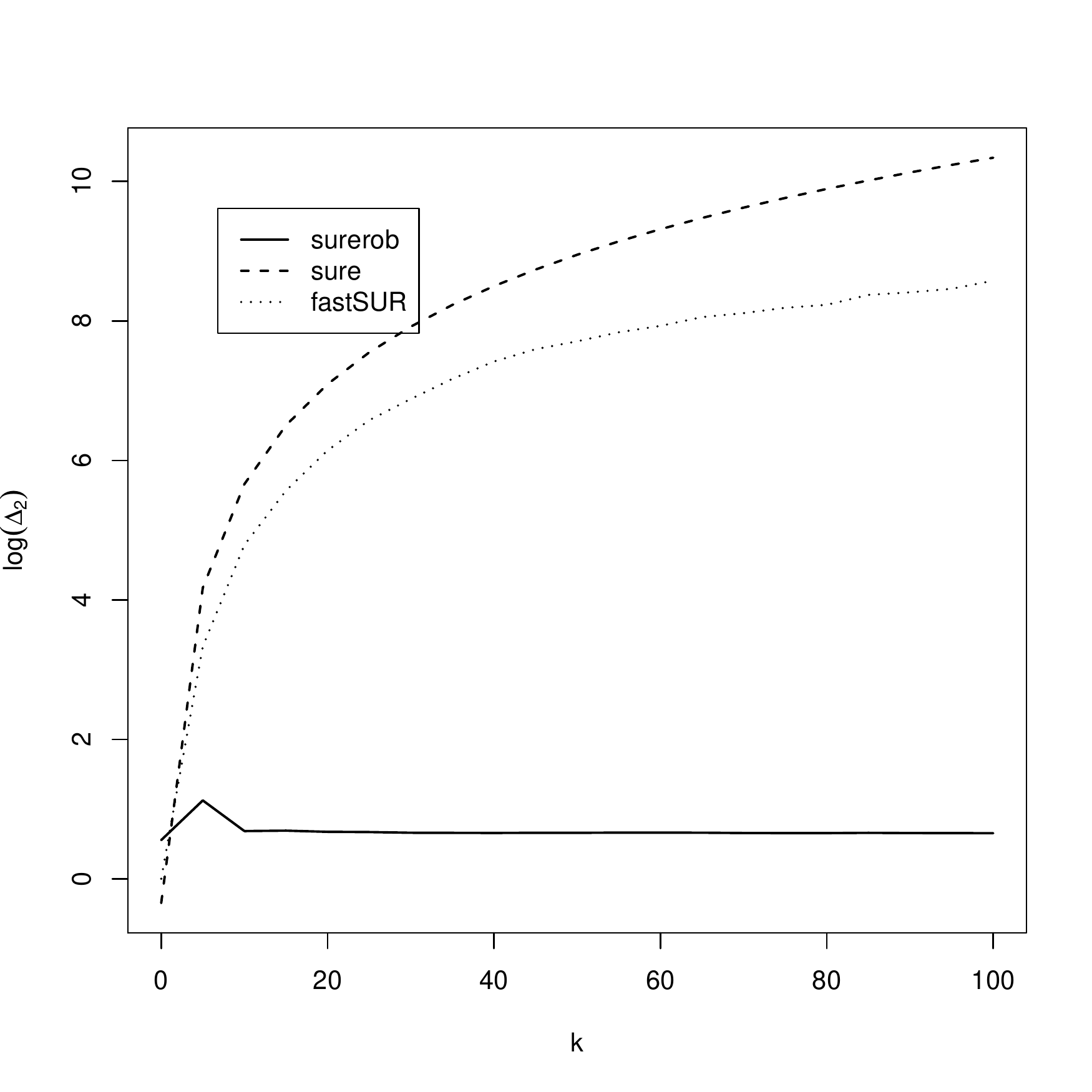}
\includegraphics[width=0.45\textwidth]{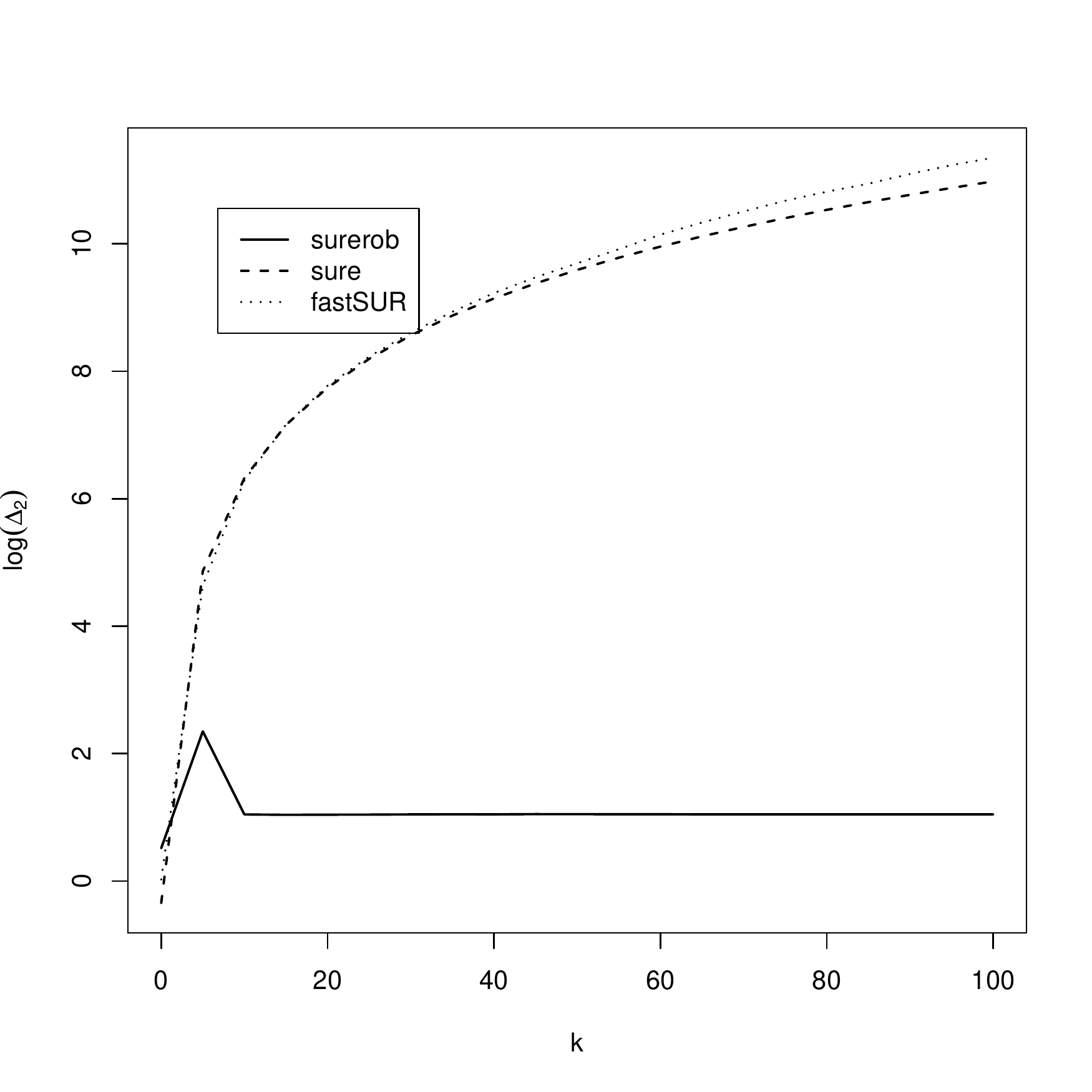} \\
\includegraphics[width=0.45\textwidth]{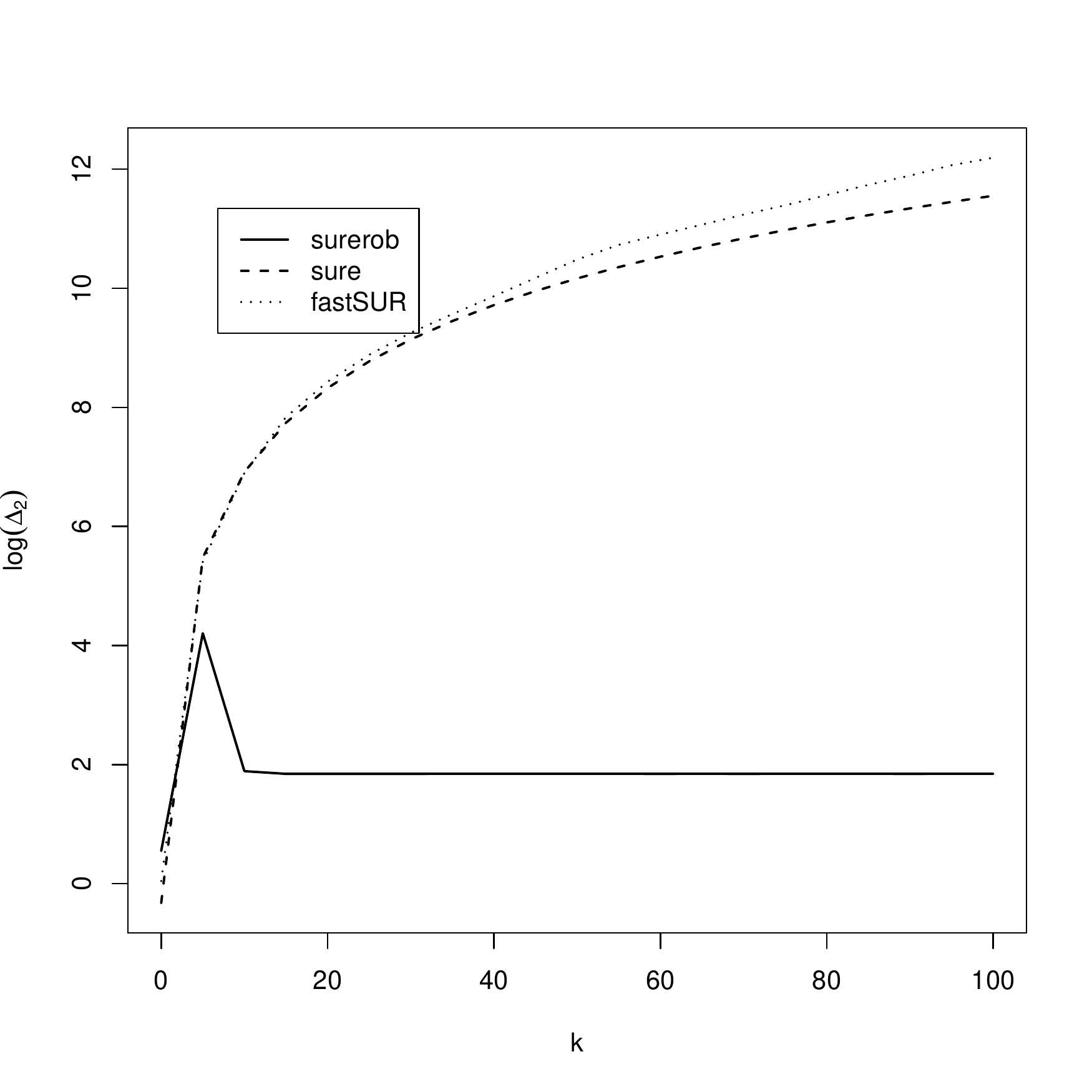}
\includegraphics[width=0.45\textwidth]{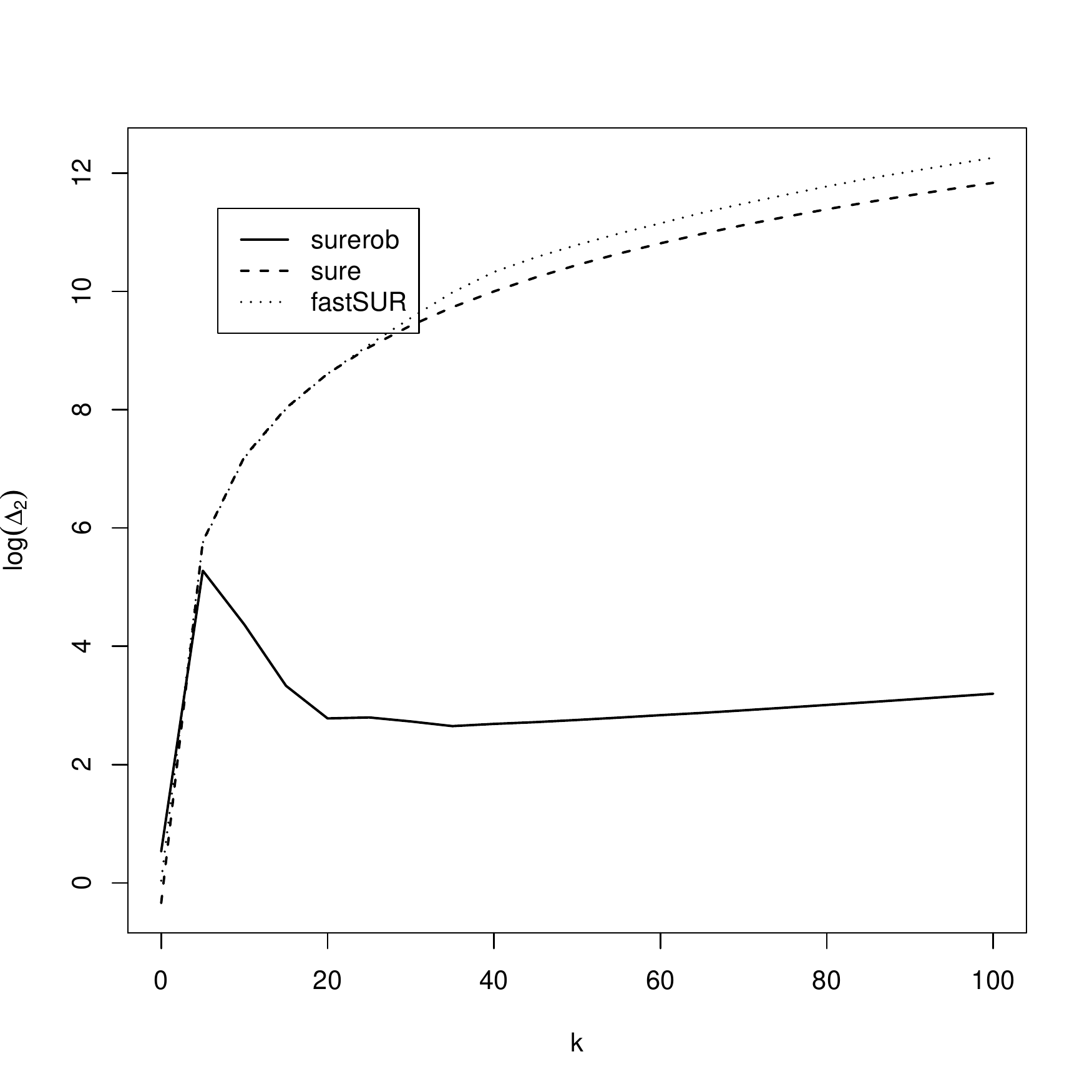}
\end{center}
\caption{Average Kullback-Leibler $\Delta_2 = \Delta(\hat{\Sigma}_2, \Sigma)$ for surerob (solid line), sure (dashed line) and fastSUR (dotted line) for different levels of contamination $\epsilon=5\%$, $10\%$, (top) $20\%$, $30\%$ (bottom) under the ICM considering $p=5$ and number of equations $m=10$.}
\label{fig:div2:100:icm}
\end{figure}

Figures \ref{fig:mse:100:htcm}-\ref{fig:div2:100:htcm} report the results for the case $p=5$, $m=10$ for THCM, while Figures \ref{fig:mse:100:icm}-\ref{fig:div2:100:icm} report the results for ICM. Results are similar for all the other cases and they are not reported. Complete results are available in Section SM--2 of the Supplemental Material. As expected, the sure method is sensitive to the presence of contamination. The fastSUR has a better performance than surerob only for the THCM with low level of contamination, says under $10\%$, while in all other cases the procedure breaks down and often performs slightly worse than the classical sure procedure. When ICM is considered surerob outperforms the other estimators.

\begin{table}[htbp]
 \caption{Mean execution time (seconds) for the three methods under different contamination type and levels, for $p=5$ and $m=10$.}
\label{tab:time}
\begin{center}
\begin{tabular}{llrrrrr}
\hline
       &      & \multicolumn{5}{c}{Contamination level} \\
Method & Type & $0\%$ & $5\%$ &  $10\%$ & $20\%$ & $30\%$ \\
\hline
surerob & THCM  &  5.89 &  5.83 &  5.74 &  5.62 &  5.82 \\
        & ICM   &  -- &  6.82 &  7.68 &  9.62 & 10.87 \\
\hline
sure    & THCM   & 0.36 &  0.36 &  0.35 &  0.36 &  0.35 \\
        & ICM    & -- &  0.36 &  0.36 &  0.36 &  0.35 \\
\hline
fastSUR & THCM  & 217.61 & 218.84 & 220.04 & 223.99 & 228.82 \\
        & ICM   & -- & 222.58 & 218.75 & 217.29 & 212.98 \\
\hline
\end{tabular}
\end{center}
\end{table}

Table \ref{tab:time} reports the mean execution time, in seconds, for the three procedures for $p=5$ and $m=10$. While the classic sure method is the fastest in all the cases, our procedure is very competitive, instead the fastSUR is by far the slowest procedure. We tried to extend the simulation study to the case $n = 500$, but the computational time needed by the fastSUR method was prohibitive.

\section{Real data example}
\label{sec:example}

In this section the introduced robust method is performed on data about tourism and compared with the results obtained with the sure method, from the \texttt{R} package \texttt{systemfit}, and the fastSUR algorithm.

\citet{Disegna2016} collected data to study the relationship between satisfaction and tourism expenditure, as well as the dependence among different tourism expenditure categories. In particular, the aim was twofold: to investigate the influence on tourism expenditure of tourists' satisfaction with the destination, considering some socio-demographic and trip-related variables; second, to study the dependence among tourist expenditure on different categories. 
Data were collected through a survey conducted by the Bank of Italy (Banca d'Italia) which includes socio-demografic characteristics, information on the trip, information on the expenditure, level of satisfaction with different aspects of the trip and overall satisfaction with the destination. In their study \citet{Disegna2016} focused on 1030 foreign visitors who visited the provinces of Bolzano, Trento and Belluno in 2011 with the main purpose of trip being ``tourism'', ``holiday'' or ``leisure''. For each observation, we have detailed information on the money spent with respect to five categories: ``Accommodation'', ``Food and Beverages'', ``International transportation'', ``Shopping'' and ``Other services'' (such as museum, excursions, shows and so on). For a complete description of the survey and the descriptive analysis of the data set, see \citet{Disegna2016}.

We considered four regression equations, one for each expenditure category excluding the ``Other services'' category. The explanatory variables considered include the satisfaction with respect to 10 classes, destination, number of nights, age and if the person is traveling alone, for a total of 20 explanatory variables. Each equation shares the same covariates except for one which express the total tourism expenditure in the remaining categories (the ``Other services'' category is included in these summations). 

\begin{figure}
\begin{center}
\includegraphics[width=0.48\textwidth]{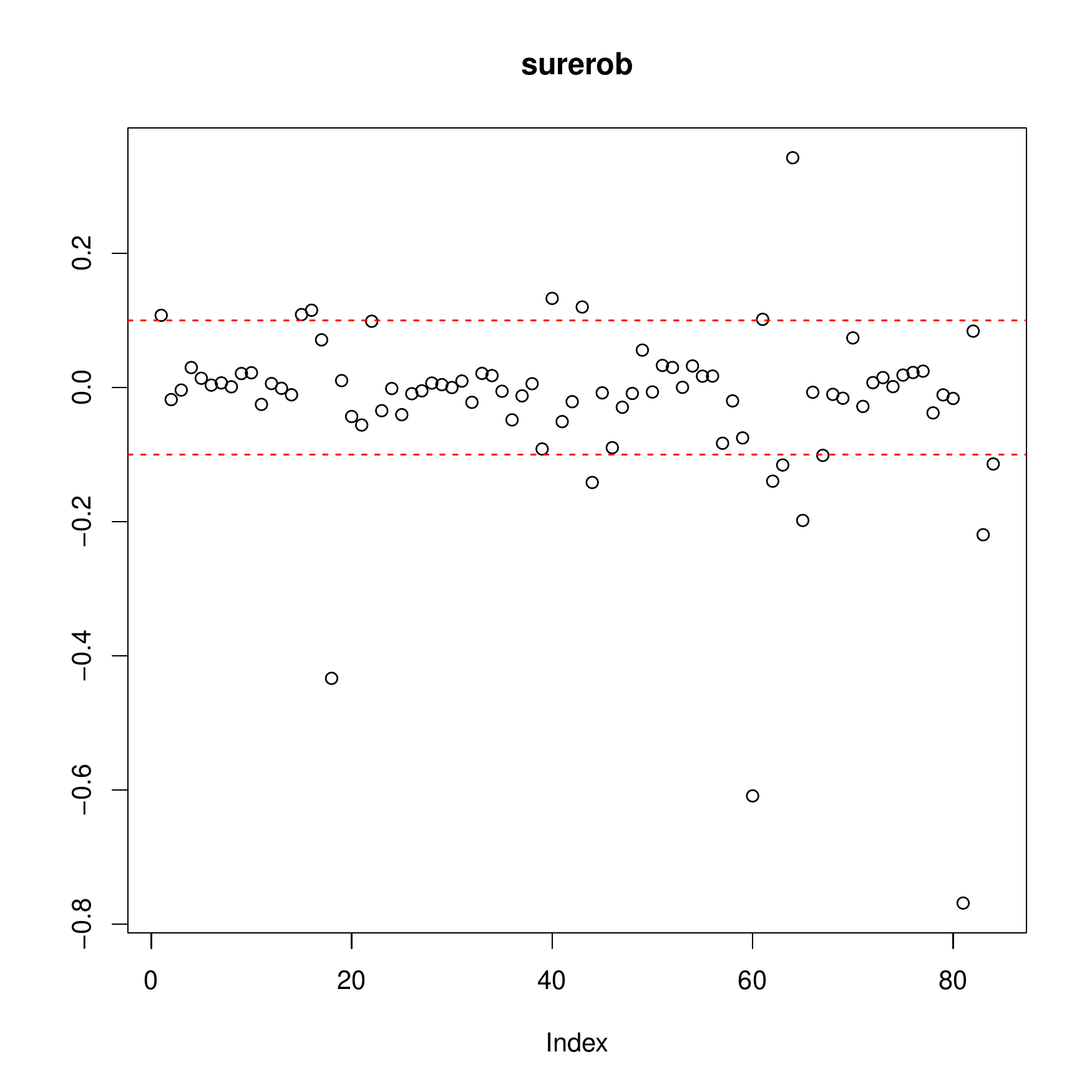}
\includegraphics[width=0.48\textwidth]{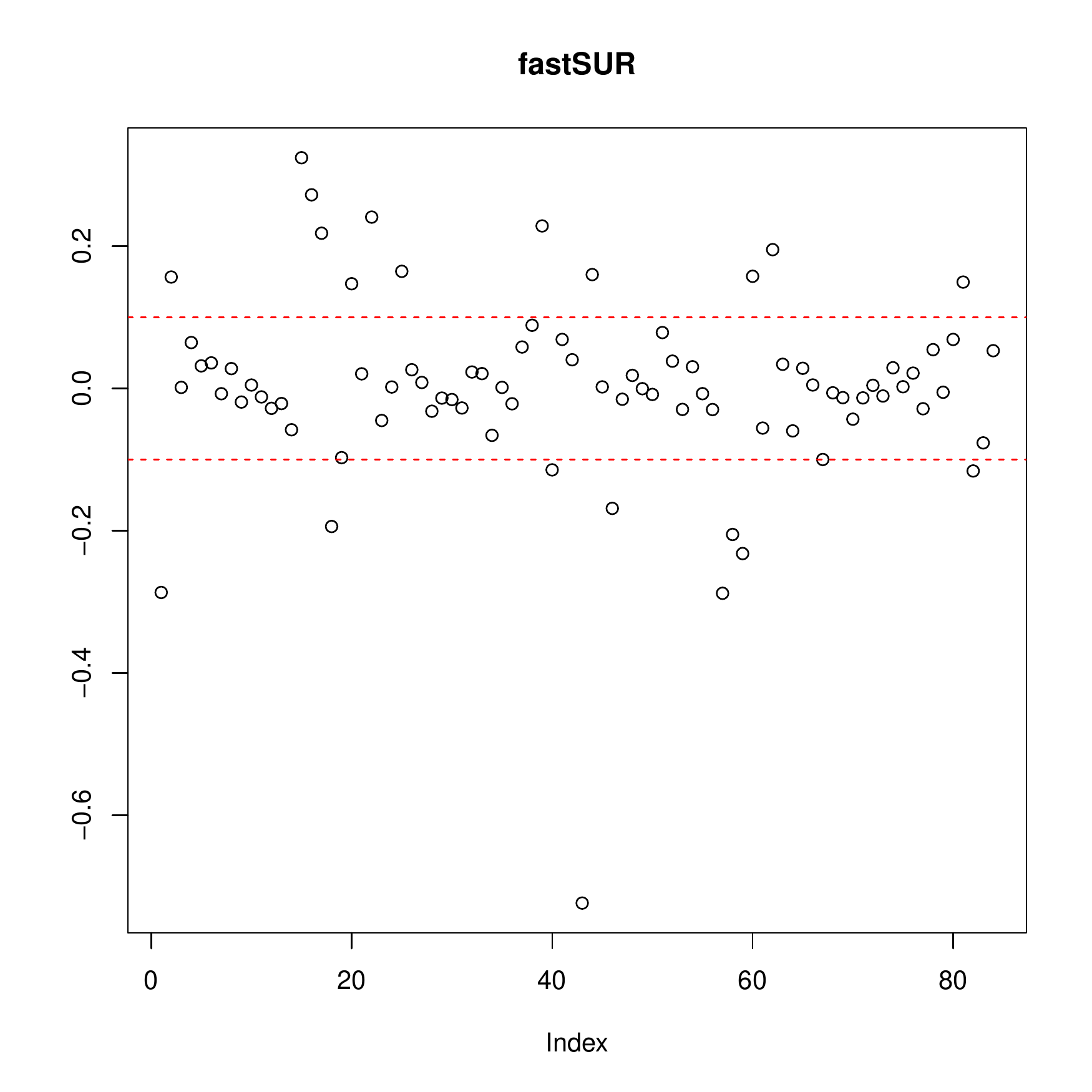}
\end{center}
\caption{On the left, differences between the estimates obtained by surerob and the estimates obtained  by sure. On the right, differences between estimates given by fastSUR and sure methods.}  
\label{fig:diff}
\end{figure}

\begin{figure}
\begin{center}
\includegraphics[width=0.5\textwidth]{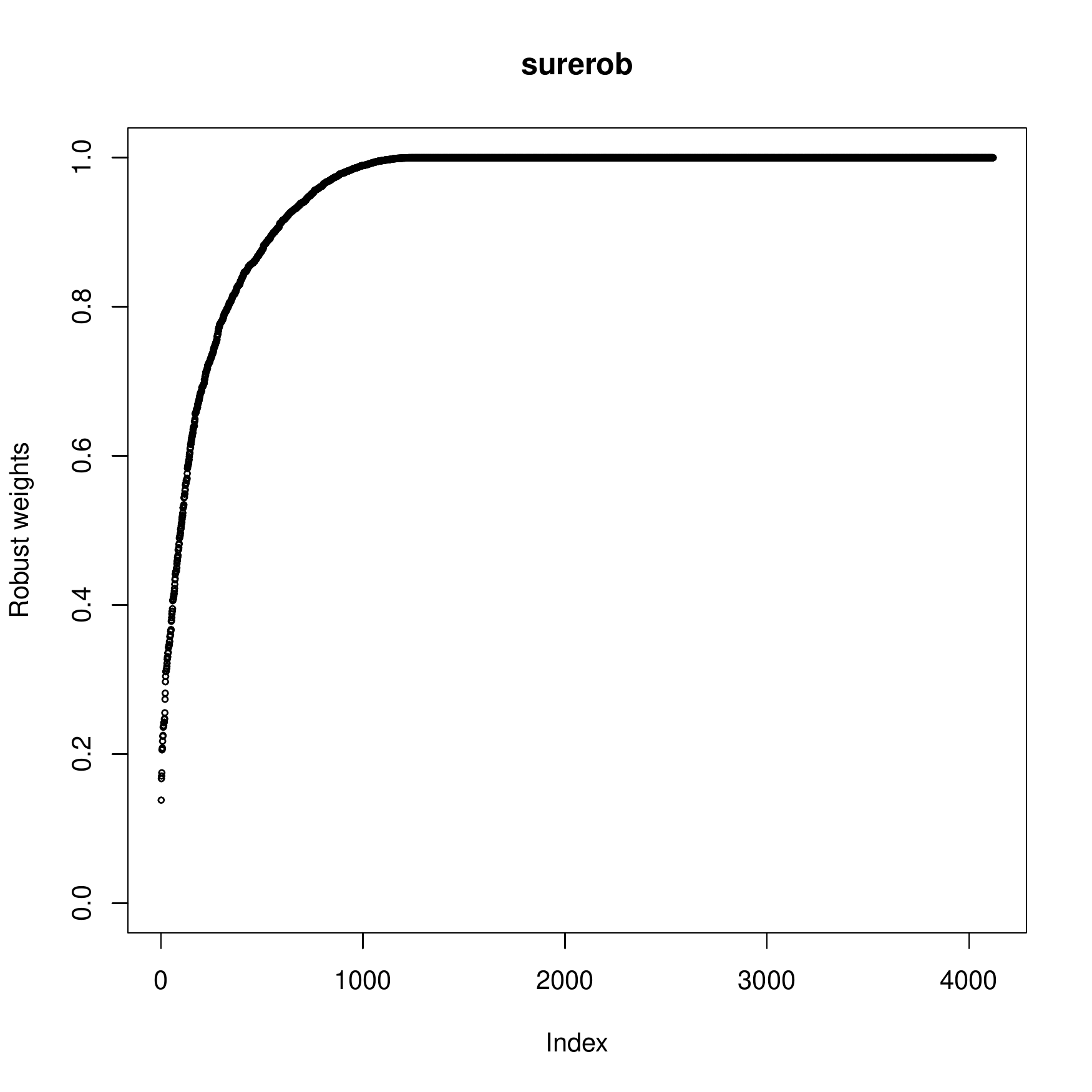}
\end{center}
\caption{Robust weights assigned to observations for each equation by the surerob method.}
\label{fig:rweights}
\end{figure}

Figure \ref{fig:diff} shows the differences between the estimates obtained by the surerob and sure, on the left, and the differences between estimates given by fastSUR and sure, on the right. We added dashed red lines highlighting the interval $(-0.1,0.1)$ on the $y$-axis. Parameters estimated by surerob are quite similar to those given by the traditional sure, with some exceptions, while the differences computed with respect to the fastSUR method show more variability.

The surerob estimation procedure is constructed assigning cell-wise weights, defined in equation \ref{eq:weights}, associated to observations for each equation, which are displayed in Figure \ref{fig:rweights}. Weights have been sorted to make the amount of downweighted observations more clear. Assume that we flag a cell as outliers if the corresponding weight is less than 0.5. In this case, surerob identifies $2.3\%$ of cell-wise contamination that propagates to the $8.8\%$ of rows. 

Finally, we computed $p$-values and $R^2$ for the estimates obtained by sure and surerob, with respect to the whole system and single regression equations. The fastSUR method has not been considered since this part has not been implemented.

\begin{table}[htbp]
  \caption{P-values computed using the estimates given by sure and surerob methods for the single regression equations.}
\label{tab:pvalues}
  \begin{center}
     \resizebox{\textwidth}{!}{
\begin{tabular}{r|rr|rr|rr|rr|}
  
  & \multicolumn{2}{c|}{Eq.1 (Accommodation)} & \multicolumn{2}{c|}{Eq.2 (Transportation)} & \multicolumn{2}{c|}{Eq.3 (Food)} & \multicolumn{2}{c|}{Eq.4 (Shopping)} \\
 & sure & surerob & sure & surerob & sure & surerob & sure & surerob \\ 
  \hline
(Intercept) & 0.265 & 0.323 & 0.309 & 0.168 & \textbf{0.003} & \textbf{0.003} & 0.110 & 0.264 \\ 
  luoghi\_visitati1 & 0.870 & 0.664 & \textbf{0.000} & \textbf{0.000} & 0.109 & \textbf{0.005} & 0.073 & \textbf{0.000} \\ 
  nr\_notti & \textbf{0.007} & \textbf{0.000} & \textbf{0.001} & \textbf{0.000} & 0.062 & \textbf{0.000} & 0.171 & \textbf{0.000} \\ 
  viaggia\_solo\_971 & \textbf{0.026} & \textbf{0.024} & \textbf{0.000} & \textbf{0.000} & 0.084 & 0.458 & 0.654 & 0.103 \\ 
  voto\_cortesia & 0.674 & 0.334 & 0.141 & 0.058 & 0.057 & 0.324 & 0.702 & 0.433 \\ 
  voto\_arte & 0.769 & 0.807 & 0.948 & 0.834 & 0.731 & 0.910 & 0.604 & 0.814 \\ 
  voto\_ambiente & \textbf{0.001} & \textbf{0.000} & \textbf{0.002} & \textbf{0.000} & 0.872 & 0.164 & \textbf{0.013} & 0.225 \\ 
  voto\_alberghi & 0.760 & 0.738 & \textbf{0.027} & \textbf{0.023} & 0.150 & 0.144 & 0.556 & 0.984 \\ 
  voto\_pasti & 0.398 & 0.098 & \textbf{0.014} & \textbf{0.008} & 0.214 & 0.699 & 0.559 & 0.582 \\ 
  voto\_prezzi & \textbf{0.046} & \textbf{0.001} & 0.07640 & \textbf{0.019} & 0.116 & 0.699 & 0.660 & 0.246 \\ 
  voto\_acquisti & 0.205 & \textbf{0.023} & 0.937 & 0.607 & 0.257 & 0.308 & \textbf{0.004} & \textbf{0.000} \\ 
  voto\_informazioni & 0.419 & 0.251 & 0.375 & 0.112 & 0.754 & 0.627 & 0.551 & 0.201 \\ 
  voto\_sicurezza & 0.292 & 0.250 & 0.051 & 0.091 & 0.714 & 0.991 & 0.089 & \textbf{0.007} \\ 
  voto\_complessivo & 0.480 & 0.546 & 0.229 & 0.156 & 0.467 & 0.354 & 0.689 & 0.963 \\ 
  eta\_cod11 & 0.140 & 0.461 & 0.115 & \textbf{0.034} & 0.427 & 0.881 & \textbf{0.046} & \textbf{0.026} \\ 
  eta\_cod21 & 0.167 & 0.604 & 0.736 & 0.634 & 0.492 & 0.638 & 0.352 & 0.279 \\ 
  eta\_cod31 & \textbf{0.028} & \textbf{0.049} & 0.742 & 0.687 & 0.683 & 0.906 & 0.355 & 0.296 \\ 
  stato\_21 & \textbf{0.000} & \textbf{0.000} & \textbf{0.000} & \textbf{0.000} & 0.079 & \textbf{0.000} & \textbf{0.000} & 0.714 \\ 
  stato\_31  & 0.420 & 0.289 & 0.236 & 0.783 & 0.315 & 0.782 & 0.063 & \textbf{0.003} \\ 
  stato\_41 & \textbf{0.013} & \textbf{0.020} & 0.821 & 0.362 & 0.096 & 0.978 & 0.121 & 0.440 \\ 
  total\_sum & \textbf{0.000} & \textbf{0.000} & \textbf{0.000} & \textbf{0.000} & \textbf{0.000} & \textbf{0.000} & \textbf{0.000} & \textbf{0.000} \\ 
   \hline
\end{tabular}
}
\end{center}
\end{table}

\begin{table}[htbp]
 \caption{R-squared computed using the estimates given by sure and surerob methods for the single regression equations and for the system.}
\label{tab:r-squared}
\begin{center}
\begin{tabular}{l|rrrr}
  & \multicolumn{2}{c}{sure} & \multicolumn{2}{c}{surerob}\\
 & $R^2$ & adj $R^2$ & $R^2$ & adj $R^2$ \\ 
  \hline
  Eq.1 (Accommodation) & 0.465 & 0.454 & 0.527 & 0.518 \\ 
  Eq.2 (Transportation) & 0.340 & 0.331 & 0.372 & 0.359 \\ 
  Eq.3 (Food) & 0.344 & 0.331 & 0.378 & 0.366 \\ 
  Eq.4 (Shopping) & 0.154 & 0.137 & 0.188 & 0.172 \\ 
  System & 0.325 & - & 0.457 & - \\
  \hline
\end{tabular}
\end{center}
\end{table}

Table \ref{tab:pvalues} reports the $p$-values for each covariate considering the equations separately. Remember that the last variables is the only one that differs among equations. Considering a $95\%$ confidence level, significant $p$-values are written in bold font. Table \ref{tab:r-squared} reports the $R^2$ values computed for sure and surerob with respect to the whole system and single equations. Robust estimates lead to a set of significant variables slightly different from that identified by standard sure, indeed some variables with a large $p$-value for sure become significant for surerob and vice versa.  

\section{Conclusions}
\label{sec:conclusions}

We proposed a new robust estimation method for the SUR model considering both row-wise and cell-wise outliers. Under the THCM, our estimator outperforms the robust competitor for high contamination levels and it remains competitive for low levels of contamination. It is worth remarking that, even if it is slower than the classic sure method, the surerob method is less computational expensive than the other robust estimators. This is an appealing property when $n$ increases. When cell-wise contamination is considered, the proposed estimator shows the best performance.

\bibliography{rsur}

\end{document}